# Diamond Diode for Extreme Venus Environments

Harshad Surdi, Gabriel Munro-Ludders[1], Mohamadali Malakoutian[2], Srabanti Chowdhury[2], Savannah Eisner, Mohamed Fadil Isamotu, Mengyang Yuan, Stephen Goodnick[1], Franz Koeck[1], James Lyons[1], Trevor Thornton[1], and Robert Nemanich

***Abstract*— A diamond Schottky PIN diode (SPIND) with the highest reported current density to date of $\sim 116\ \mathrm{kA/cm^2}$ is demonstrated carrying a total current of $\sim 1.3\ \mathrm{A}$ through a 50 $\mu m$ wide pseudo-vertical diode structure. The diamond SPIND also provides a maximum power handling capacity of $1.85\ \mathrm{MW/cm^2}$ and a low specific on-resistance $R_{on}S$ of $0.05\ \mathrm{m\Omega.cm^2}$ at a forward bias of $\sim 16\ \mathrm{V}$. The diamond SPIN diode also shows excellent rectification characteristics with a current on-off ratio of $\sim 6 \times 10^{12}$. An analytical model including thermionic emission and space charge limited current is presented together with Silvaco ATLAS TCAD simulations, to accurately reproduce the experimental J-V characteristics using multiple single trap levels and other physical models emulating a real device. Theoretical analysis from the analytical models in conjunction with ATLAS simulations shows that further improvement in the device turn on voltage and $R_{on}S$ can be achieved by reducing the defect density and contact resistance in order to approach the ultimate performance in the Mott-Gurney space charge limited current regime.**

## 1. Introduction

Venus and Earth share similar size and density, yet past missions have revealed Venus to be an extreme environment—hot, dry, lacking plate tectonics, and enveloped in a dense, reactive atmosphere. Understanding Venus's atmospheric evolution, interior–surface interactions, and potential past habitability remains a major priority for upcoming NASA missions. Achieving these science goals requires long-duration surface operations, including seismic measurements, mineralogical characterization, and monitoring of atmospheric chemistry over extended periods.

Historically, surface missions have been severely constrained by electronics survivability. Although ten Soviet landers reached the surface, the longest operation—Venera-13—lasted just over two hours before silicon electronics failed in the ~460 °C, ~92-bar environment. Future missions demand systems capable of >60 days of continuous operation at ≥500 °C, a requirement that exceeds the limits of conventional semiconductor technologies.

High-temperature diamond electronics directly address this challenge. Diamond's wide bandgap, extremely low intrinsic carrier concentration, high thermal conductivity, and chemical inertness make it uniquely suited for stable operation in Venus-like conditions. These capabilities align with the objectives of NASA's HOTTech program, which seeks to reduce risk for future high-temperature missions, including Venus landers, Mercury surface probes, and deep atmospheric probes for gas giants.

While prior work has demonstrated diamond diodes operating at temperatures up to 1000 °C, long-duration functionality also requires advances in high-temperature packaging, interconnects, and test methodologies. Developing these technologies is essential for enabling robust diamond-based power electronics that can support sustained scientific exploration in extreme planetary environments.

This work demonstrates fully operational diamond diodes in simulated Venusian atmosphere for 15 days under constant electrical stress.

## 2. Growth and Device Fabrication

Phosphorus-doped n-layers (~50 nm) and intrinsic i-layers (~300 nm) were grown on heavily boron-doped p-type <111> HPHT diamond substrates (3×3 mm, $\sim 3\times10^{20}\ \mathrm{cm^{-3}}$, 300 µm thick), forming the p-i-n diode stack. The n-layer doping is $\sim 3\times10^{17}\ \mathrm{cm^{-3}}$, while the intrinsic layer exhibits background doping on the order of $10^{14}\ \mathrm{cm^{-3}}$. Both layers were deposited by PECVD, with growth details described in [7]. Circular diodes (50 µm diameter) were defined by partially etching the intrinsic layer using $O_2/SF_6$ RIE with a $SiO_2$ hard mask, creating isolated mesas. Cathode contacts (38 µm diameter) were patterned on the n-layer using a Ti/Pt/Au stack (50/50/300 nm). Anode contacts to the $p^{++}$ substrate were formed by etching trenches adjacent to the mesa and depositing the same metal stack, spaced 90 µm from the cathode. Owing to the near-metallic boron doping, the metal stack forms a Schottky cathode and an ohmic anode. Figure 1(a) illustrates the pseudo-vertical SPIN diode structure, and Figure 1(b) shows the fabricated device.

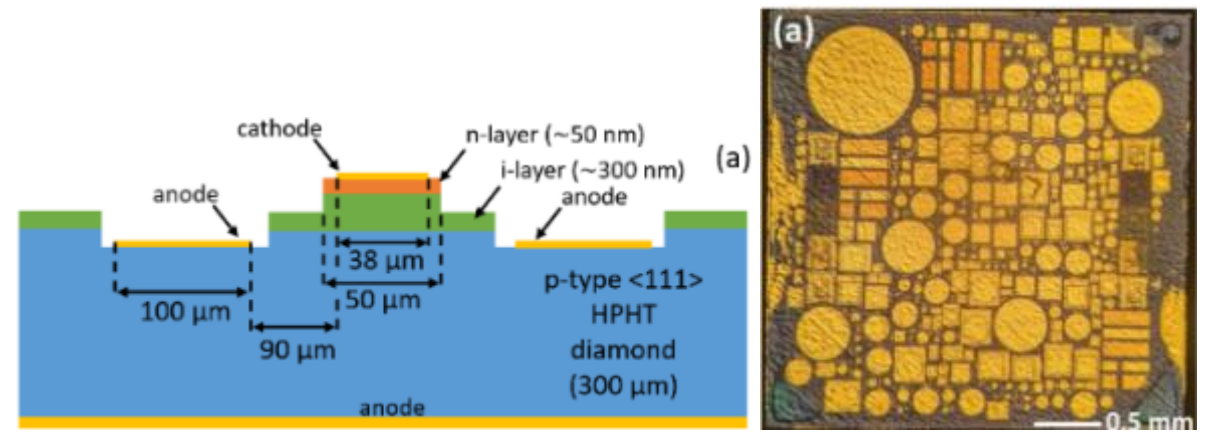

*Fig. 1: (a) Cross-section of diamond SPIN diode (b) Fabricated diamond SPIN diode device*

## 3. Experimental Setup

### A. Pre-GEER IV Measurements

High-temperature performance of the patterned GEER test devices was measured from room temperature to 500 °C using a computer-controlled Linkam THMS600 hot stage. Figure 2 shows the logarithmic and linear J–V characteristics of sample 17-050 across this temperature range, with detailed forward- and reverse-bias analysis provided in the following sections. The diode delivers a forward current density of ~20 A/cm² at ~10 V at 25 °C and at ~5.5 V at 500 °C. The turn-on voltage decreases from ~7.9 V at room temperature to ~4.7 V at 500 °C. The ideality factor increases non-linearly from ~2.2 at room temperature—consistent with near-ohmic contacts and two-carrier conduction—to ~3.8 at 500 °C, indicating growing non-idealities such as interface states, barrier inhomogeneity, and image-force effects. A sharp drop in ideality factor to ~2.1 at 225 °C suggests contact annealing during thermal cycling.

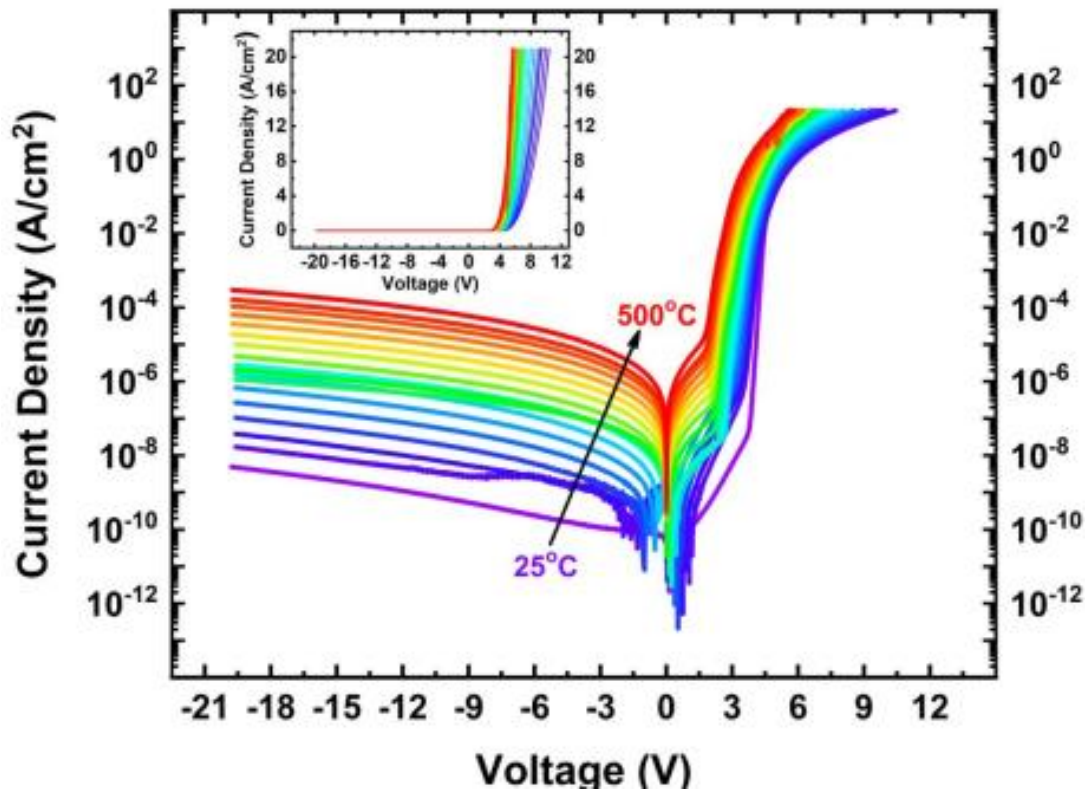

*Fig. 2 Pre-GEER temperature cycling IV*

### B. High Temperature Packaging and Data Acquisition Setup

The active device (17-050) was additionally mounted on a ceramic chip carrier, where the 800 µm-diameter diode's cathode was wire-bonded to the carrier pads (Figure 3). Ground connections were wire-bonded to the top and bottom $p^{++}$ anode contacts. All high-temperature packaging was completed in collaboration with Makel Engineering.

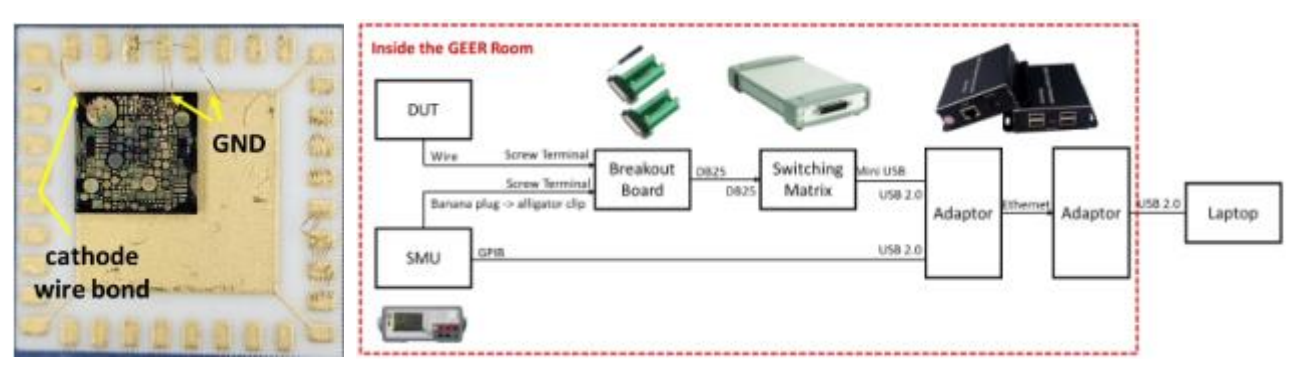

*Fig. 3 Diamond diode wire bonded to ceramic chip carrier*

The 17-050 SPIND device was selected for active in-situ characterization during the 10-day GEER test campaign. An automated data-acquisition system, developed jointly with ASU, Stanford, and MIT, enables continuous monitoring of three devices mounted on a ceramic chip carrier. The system is programmed to cycle between devices and record measurements in parallel. A schematic overview is shown in Figure 3. The 17-050 SPIND device is configured for continuous in-situ monitoring during the GEER campaign. Current is recorded at a fixed reverse bias of −10 V, and a full reverse-to-forward I–V sweep is performed every four hours. A shorted pair of gold pads on the ceramic carrier is included as a reference "resistor" to track changes in wire and interconnect resistance throughout the test.

### C. Post-GEER Measurements

Samples were characterized by XPS in a VG Scienta system (Al Kα, 1486.7 eV; $1.9\times10^{-9}$ Torr; 160 meV resolution). After GEER exposure, they were analyzed in a KRATOS Supra XPS chamber ($2\times10^{-8}$ Torr) equipped with monochromatic Al Kα, hemispherical and spherical-mirror analyzers. All spectra were acquired at ambient temperature, calibrated to the Au $4f_{7/2}$ peak (84.0 eV), and collected using a 110 µm aperture and 40 eV pass energy.

## 4. Results and Discussion

Figure 4 shows the reverse bias current vs. time for the entire GEER run. As the GEER chamber rises in temperature to full Venusian temperature and atmosphere, the reverse bias current rises and settles to ~10mA. The reverse bias current remains stabilized at 10mA till day 21 and drops back down to noise level as the GEER chamber returns to room temperature. Figure 5 shows the diode IV characteristics at idle, ramp up, at Venus atmosphere and ramp down. As expected, both the forward and reverse bias IV characteristics as the diode is sampled from idle to Venus atmosphere. The diode remains completely functional at Venus atmosphere throughout the 15 days and maintains a Ion/Ioff ratio of 10^5.

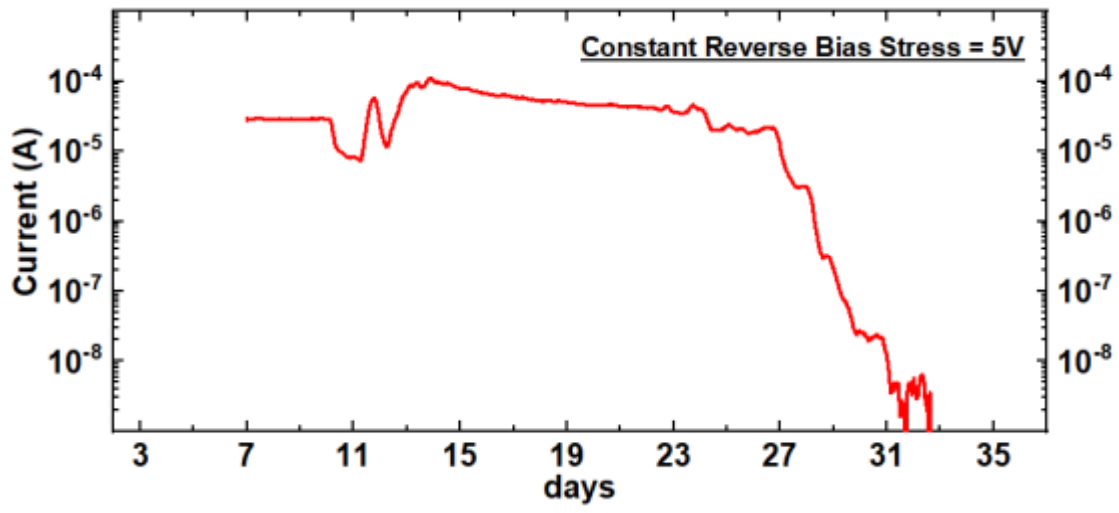

*Fig. 4 Constant reverse bias characterization thrught entire GEER run*

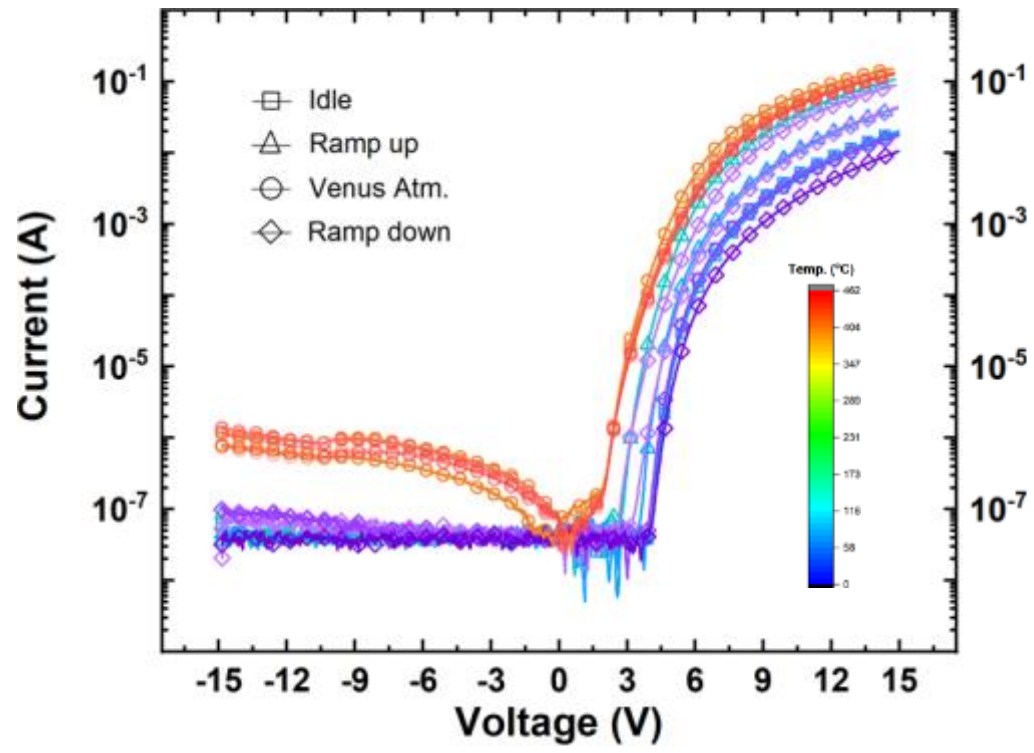

*Fig. 5 IV measurements through out entire GEER run*

XPS analysis of the un-passivated 1b surface and patterned sample 20-059 shows trace sulfur incorporation after GEER exposure, primarily as thiols, sulfuric acid, and nickel-sulfate species concentrated near the nickel contact pads. Minor increases in oxidized carbon (O–C–O, O–C=O) are also observed, while other atmospheric species remain below the <0.1 at% detection limit. Nickel oxide and sulfur-bearing compounds dominate the reacted metal surface, consistent with $SO_2$—present at the highest sulfur concentration in GEER—being the likely reactive source. Post-test spectra show a 2.2 at% rise in nickel and a reduction of gold from 11.7 at% to 0.2 at%, suggesting interlayer diffusion or partial delamination. Titanium remains undetected before and after testing, indicating that the Ti–diamond interface was not exposed to the GEER environment. Chlorine and fluorine species present in the chamber (HCl, HF) were not detected on any sample. XPS measurements show no titanium on any sample after GEER exposure, indicating that the Ti/Ni/Au contact stack remained intact and the Ti–diamond interface was not compromised. Likewise, electrical data from the active device show no degradation; instead, performance improves over the test duration, confirming that the metal contacts did not fail mechanically or electrically during the high-temperature Venus-analog environment.

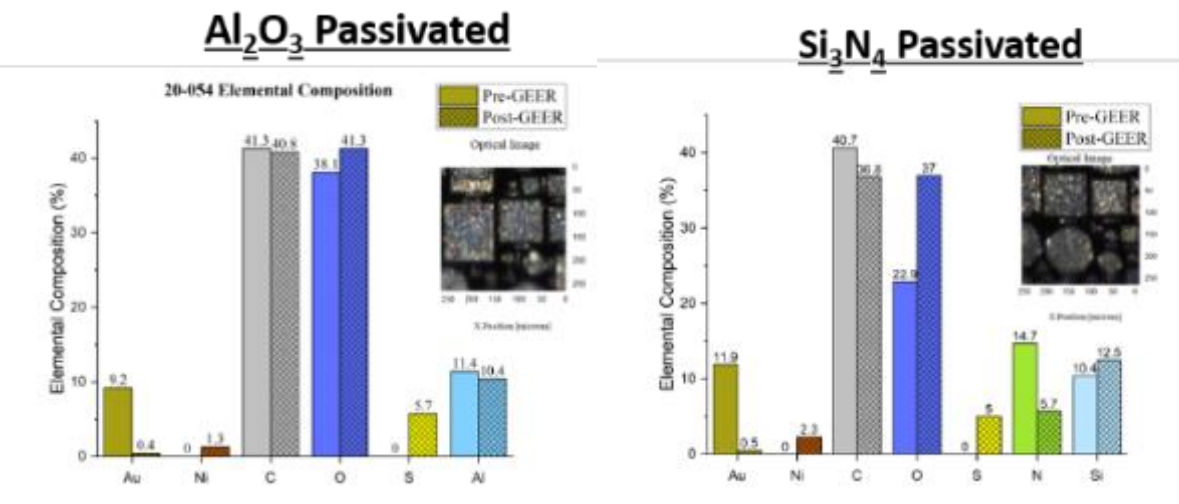

*Fig. 6 Post-GEER XPS characterization*

## 5. Conclusion

Diamond diodes have been successfully demonstrated to operate at Venus atmosphere for 15 days under constant rever bias and full IV characterization. The diodes remain fully functional maintaining Ion/Ioff ratio of 10^5 throughout the GEER run. The XPS measurements post-GEER also show minimal degradation in contacts and diamond surface. Thus the diamond devices are perfectly suited for extreme environment Venus applications.

## Acknowledgment

This work was funded under the NASA HOTTech grant no. GR02231. The authors also acknowledge and are thankful for the continued support of all staff members at ASU NanoFab, funded, in part, by NSF grant ECCS-1542160.